\documentclass[12pt]{iopart}
\usepackage{graphics,epsfig,amssymb}

\begin{document}

\title[Construction of equilibrium networks]
{Construction of equilibrium networks with an energy function}

\author{Daun Jeong\dag, M.Y. Choi\dag\ddag, Hyunggyu Park\ddag}
\address{\dag Department of Physics and Astronomy, Seoul National University, Seoul 151-747, Korea}
\address{\ddag School of Physics, Korea Institute for Advanced Study, Seoul 130-722, Korea}

\begin{abstract}
We construct equilibrium networks by introducing an energy function
depending on the degree of each node as well as the product of
neighboring degrees.
With this topological energy function, networks constitute a
canonical ensemble, which follows the Boltzmann distribution for
given temperature.  It is observed that the system undergoes a topological phase
transition from a random network to a star or a fully-connected network
as the temperature is lowered. Both mean-field analysis and
numerical simulations reveal strong first-order phase transitions
at temperatures which decrease logarithmically with the system
size. Quantitative discrepancies of the simulation results from
the mean-field prediction are discussed in view of the strong
first-order nature.
\end{abstract}
\maketitle

\section{Introduction}
\label{sec:intro}

Networks describe interaction patterns of various complex systems.
In many cases, interactions switch constantly and their
topology changes according to external conditions. If we treat a
network as a thermodynamic system which has an energy function
depending on its topology and introduce temperature as the
disorder strength, networks constitute a topological ensemble of
equilibrium networks of which physical quantities are described by methods
of statistical mechanics~\cite{ref:Vicsek,ref:Vicsek2,ref:Newman}.
At equilibrium, links are reallocated to satisfy the detailed balance
and ergodicity; topological phase transitions are expected as
the temperature is varied.
Since there does not exist the physical energy of a network associated with its topology,
the functional form of the energy is not given a priori:
It may be taken as a function of the node degree, number of links,
or some global property of the network.
If an energy function is designed for certain performance, one can obtain an
optimized network ensemble which minimizes the energy via zero-temperature dynamics.
On the other hand, various topologies can be obtained at finite temperatures, with the entropy
taken into account.
A number of networks have been constructed through the use of energy functions
with a focus on structural transitions~\cite{ref:Vicsek, ref:Manna, ref:Lassig} and
optimization~\cite{ref:Selection, ref:Optimization}.
In general, equilibrium networks constructed by a topological energy function are not complex
but random~\cite{ref:ER} at high temperatures;
they are rather simple at low temperatures as well,
making either star or fully-connected networks.
Exceptionally, scale-free networks come out only at the critical temperature
if a logarithmic function of the degree is introduced for the energy~\cite{ref:Vicsek}; 
this is logically obvious in view of that the corresponding rewiring dynamics reduces to 
the preferential attachment scheme. 

In this study, we introduce a simple function of the degree as the energy function, 
which combines two competitive terms with different strengths, and examine the resulting 
topological phase transitions.  Each term favors a different ground state, so one may
expect various network topologies depending on the temperature and the relative strength.

There are five sections in this paper: Section \ref{sec:MS} introduces the model,
together with the energy function.  In Sec. \ref{sec:MF}, the
system is analyzed by means of the mean-field theory.
Specifically, the energy and the entropy of the system are
evaluated, in terms of which phase transitions between random,
star, and fully-connected network phases are probed.
Section~\ref{sec:num} is devoted to numerical simulations. In
particular the simulation results are discussed, in comparison
with the mean-field results.  Finally, a brief summary and
concluding remarks are given in Sec.~\ref{sec:conclusion}.

\section{Model system}
\label{sec:MS}

We consider a system of $N$ nodes, some of which are connected with each other.
The total number of links is given by $M=(N/2)\langle k\rangle$, where
$\langle k\rangle$ is the mean degree, i.e., the mean number of links per node. 
In this study, $\langle k \rangle$ is fixed to be $0.5$ and $M$ is kept conserved 
during rewiring. 

Associated with the system is an energy function, which consists of two parts,
with strengths $J_1$ and $J_2$, respectively.
Specifically, we write the energy function in the form
\begin{equation}
E = -\frac{J_1}{M}\sum_{\langle i,j \rangle} k_i k_j - \frac{J_2}{M}\sum_{i}k_i^2 ,
\label{eq:Hamiltonian}
\end{equation}
where $k_i$ is the degree of node $i$ and $\langle i,j \rangle$ represents connected nodes.  
The strength of each part is normalized by the total number of links, 
so that the ground-state energy is extensive.  
This is the simplest form to describe interactions between nodes in terms of the degree, 
still affording possibility of various network topologies.  

\begin{figure}
\centerline{\epsfig{file=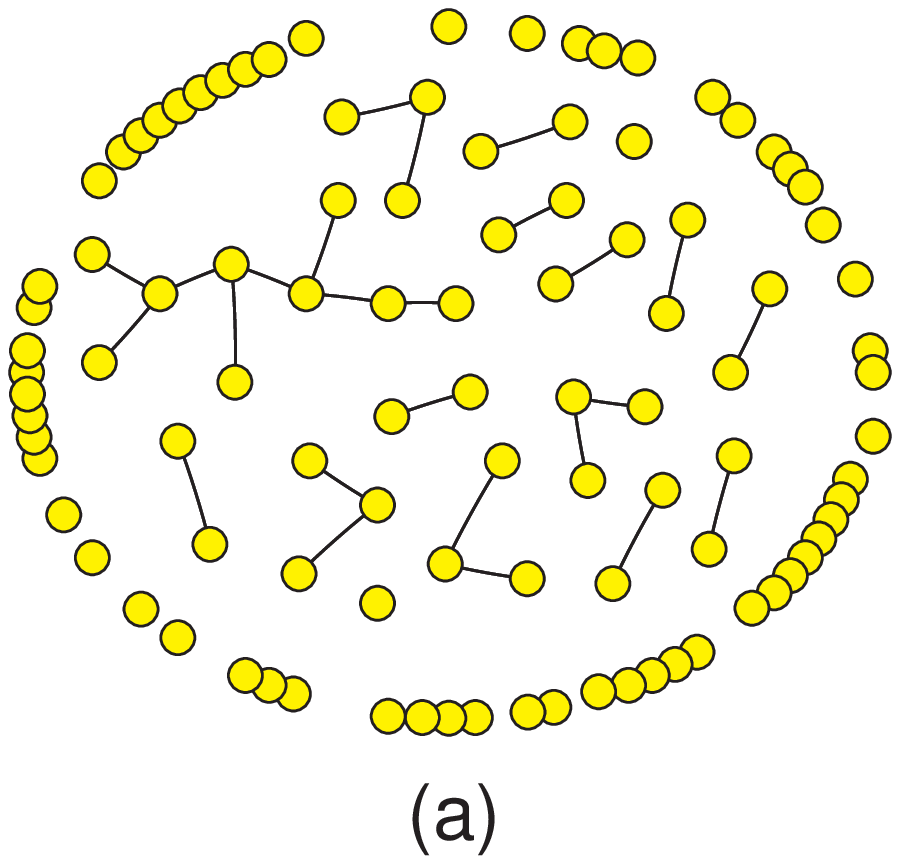,width=5cm}}
\centerline{\epsfig{file=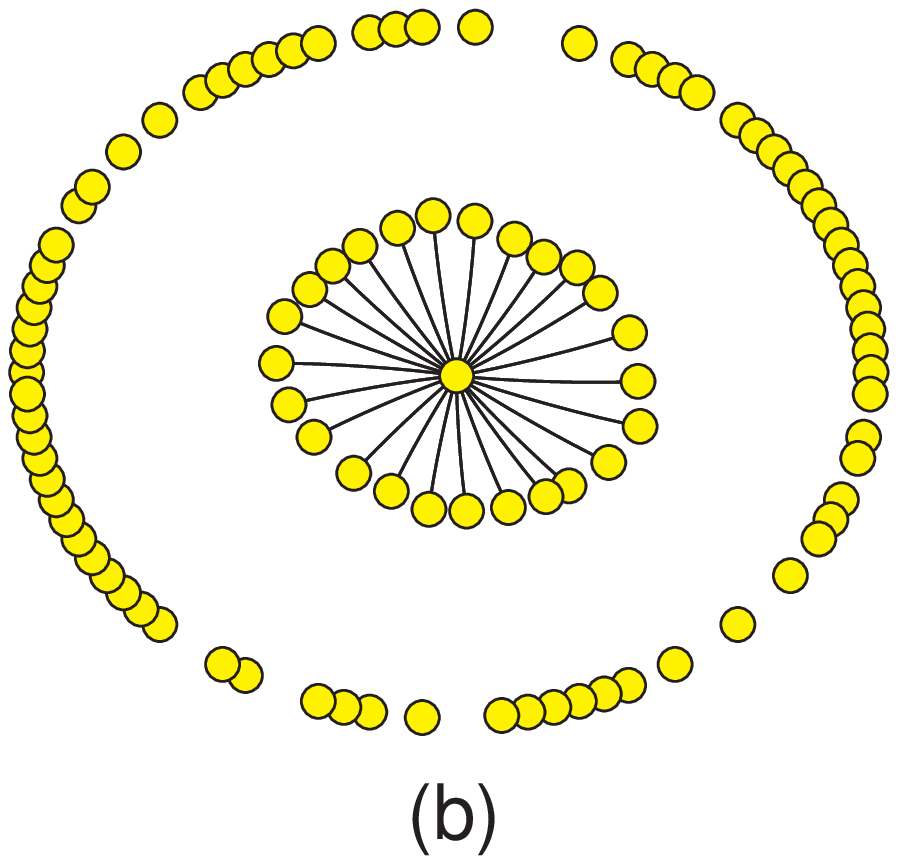,width=5cm}}
\centerline{\epsfig{file=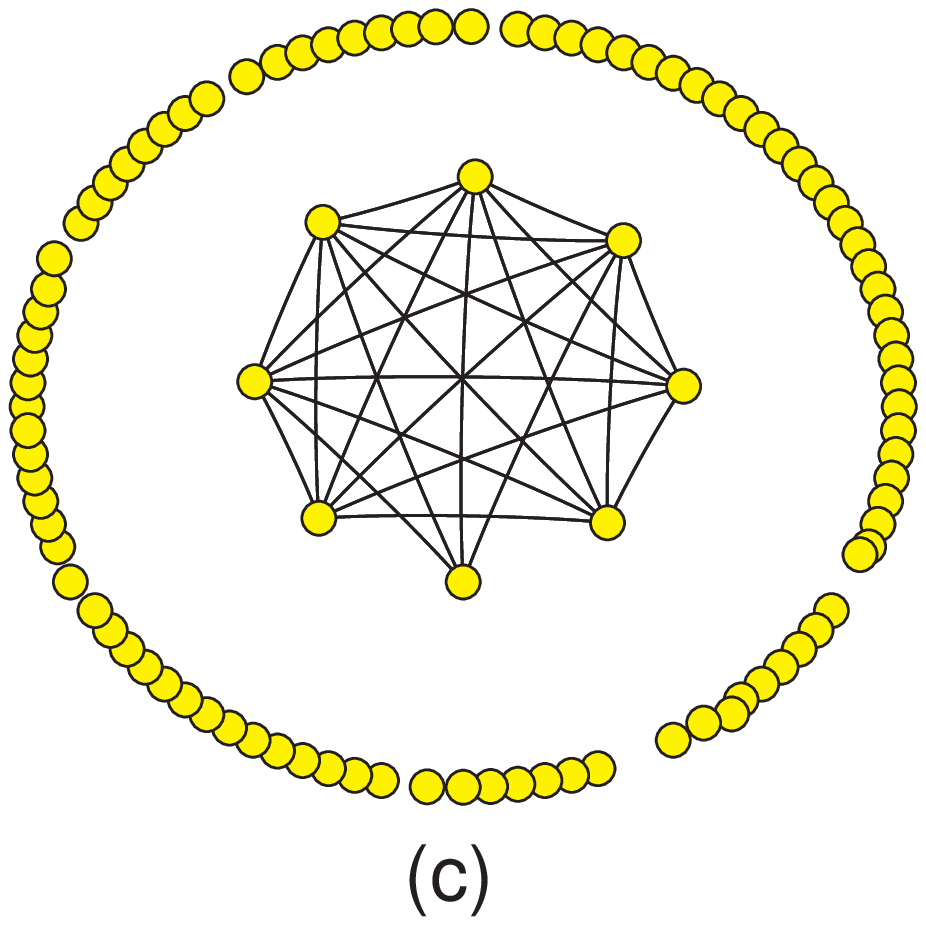,width=5cm}}
\caption{Typical configurations of (a) random, (b) star, and (c) fully-connected networks, 
generated by simulations performed on a system of $N = 100$ nodes and $M=25$ links
with $(J_1,  J_2) =$ (a) $(2, 1)$, (b) $(0.5, 1)$, and (c) $(2, 1)$
at temperature $T =$ (a) $1.36$, (b) $0.12$, and (c) $0.2$.
}
\label{fig:net}
\end{figure}

To minimize the first part, it is advantageous for a large-degree
node to be connected to another large-degree one.  In contrast,
the second part favors one node to have a degree as large as
possible.  Accordingly, the two different configurations compete
energetically: When $J_2=0$, the ground state corresponds to a
fully-connected network for given number of links. The opposite
case ($J_1=0$) yields a star network, where one node takes all
possible links. The ratio $J_1/J_2$ thus determines the ground
state to be either of both extremes or in-between. At high
temperatures, on the other hand, the system is in the disordered
state characterized by a random network, regardless of the
relative strength of the two parts.  Typical configurations of the
three types of network are shown in Fig. \ref{fig:net}.

In view of these, we expect that the system displays a topological
phase transition between a random network and one of the two
compact networks mentioned above.  We first describe the phase
transition at the mean-field level, which discloses that the phase
transition occurs only at finite system sizes and has strong
first-order nature. To confirm this, we carry out numerical
simulations and observe discontinuity in energy, hysteresis, and
metastability.

\section{Mean-field analysis}
\label{sec:MF}

We begin with the mean-field analysis of the phase transitions
between the disordered phase (the random network) and the ordered
one (either the fully-connected network or the star network). To
describe those transitions, we introduce different order
parameters, defined to become $O(1)$ in one configuration and
$O(1/M)$ in the other.  Expressed as a function of the order
parameter and the temperature, the free energy has a maximum
between the two extreme configurations due to the entropy
contributions at finite temperatures. Comparing the free energies
of the two configurations, we determine the low- and the
high-temperature phases.

\subsection{Phase transition between star and fully-connected networks}
We first consider the transition between star (S) and fully-connected (F)
networks and probe the phase boundary at which the two phases adjoin.
%
To distinguish the two, we introduce the order parameter measuring
the number of nodes having the largest degree,
\begin{equation}
\phi = \frac{n_{k_{max}}}{n}
\end{equation}
where it is assumed that among the $n$ non-isolated nodes there
exist only two kinds of nodes: $n_{k_{max}}\,(=n\phi)$ nodes of
the largest degree (fully connected to each other) and the remaining
$n(1-\phi)$ nodes connected only to the former $n_{k_{max}}$
nodes, i.e., having $n\phi$ links.  We consider an ensemble of a
fixed total number of links, $M$, which are distributed among
$n$ non-isolated nodes only. Then $n$ and $\phi$ are related via
\begin{equation}
\frac{n\phi(n\phi-1)}{2} + n \phi n (1-\phi)=M.
\end{equation}

In terms of this order parameter, the energy reads approximately
\begin{eqnarray}
E &=& -\frac{J_1}{2M}[(n-1)^2 n\phi(n\phi-1) + 2n(n-1)(n\phi)^2
(1-\phi)] \nonumber\\ 
  & & -\frac{J_2}{M}[(n-1)^2 n\phi + n(n\phi)^2 (1-\phi)],
\end{eqnarray}
which reaches the maximum between $\phi = (M+1)^{-1}$ and $\phi
\approx 1$ corresponding to a star and a fully-connected network,
respectively.
We thus compare the energy for the two kinds of network. 
In the case of a star network, we have $n-1=M$, and
\begin{eqnarray}\label{star}
 E_{S} &\equiv& E(\phi{=}n^{-1}) 
                = -\frac{J_1}{M}(n-1)^2 -\frac{J_2}{M} [(n-1)^2+(n-1)] \nonumber\\
                &=& -(J_1+J_2)M-J_2 .
\end{eqnarray}
For a fully-connected network, on the other hand, we have
$n(n-1)=2M$, and
\begin{eqnarray}\label{fully}
E_{F} &\equiv& E(\phi{=}1)  
           = -\frac{J_1}{M}\frac{n(n-1)^3}{2}-\frac{J_2}{M} n(n-1)^2 \nonumber\\
           &\approx& -2J_1M-2J_2\sqrt{2M} .
\end{eqnarray}
Comparison of Eqs. (\ref{star}) and (\ref{fully}) shows that
$E_{F}$ becomes lower than $E_{S}$ for $J_1/J_2 > 1-2\sqrt{2/M}+1/M$.
It is thus concluded that the ground state
corresponds to a fully-connected/star network for $J_1$
larger/smaller than $J_2$ in the thermodynamic limit ($M\rightarrow\infty$).

We next consider the phase boundary at finite temperatures. To
obtain the free energy at finite temperatures, we should take into
account the entropy, defined to be $S(\phi)\equiv \ln
\Omega_{\phi}$.  As a non-isolated node belongs to either a group
of $n\phi$ nodes of the largest degree or the other group of
remaining nodes, we write the number of distinct accessible
configurations in the form
\begin{equation}
\Omega_{\phi}=\left(\begin{array}{c}N \\ n \end{array}\right)
\left(\begin{array}{c}n \\ n\phi \end{array}\right).
\end{equation}
The first-order transition between star and fully-connected
networks is manifested by the barrier in the free energy
$F(\phi)=E-TS$ between two competing minima. As addressed, the
transition temperature may be determined by comparing the free
energy for the two phases. For a star network, we have $n\phi =1$
and $n=M+1$, thus
\begin{eqnarray}
S_{S} = \ln\left(\begin{array}{c}N\\ M+1\end{array}\right)+\ln (M+1) 
   \approx g(\alpha)M,
\end{eqnarray}
where $\alpha \equiv 2/\langle k\rangle$ and
$g(\alpha)=\ln[\alpha^{\alpha}(\alpha-1)^{1-\alpha}]$. For a
fully-connected network, $\phi=1$ and the number of non-isolated
nodes is given by $n \approx \sqrt{2M}$, which leads to the entropy
\begin{eqnarray}
S_{F}=\ln\left(\begin{array}{c}N \\ \sqrt{2M} \end{array}\right) 
 \approx \sqrt{\frac{M}{2}}\left(\ln M +2\ln\alpha+1 -\ln 2 \right).
\end{eqnarray}

The free energy for each type of network thus reads
\begin{eqnarray}
F_{S}&=&-(J_1+J_2)M-J_2 -Tg(\alpha)M \nonumber\\
F_{F}&=& -2J_1M-2J_2\sqrt{2M} 
                  - T\sqrt{\frac{M}{2}}\left(\ln M +2\ln\alpha+1 -\ln 2 \right).
\end{eqnarray}
The condition $F_{S} = F_{F}$ at $T=T_c$ then leads to the
transition temperature
\begin{equation}
T_c \approx \frac{1}{g(\alpha)} \left[J_1 - J_2 \left(1- 2\sqrt{\frac{2}{M}}+\frac{1}{M}\right)\right],
\end{equation}
below which the system makes a fully-connected network. Note that
the transition exists only when $J_1>J_2$ in the thermodynamic
limit. This analysis does not include the true high-temperature
phase, which is described by the random network; this is
considered in the following section.

\subsection{Phase transition between random and fully-connected networks}
To examine the phase transition between a random (R) network and a fully-connected one,
we define the order parameter $\psi$ to be the average connectivity $k$ of
non-isolated nodes relative to the number $n$ of such nodes:
\begin{equation}
\psi=\frac{k}{n},
\end{equation}
where the number of total links $M=nk/2$ is constant, thus leading to $\psi = 2M/n^2$.
The order parameter $\psi$ defined above conveniently characterizes the two phases,
taking the values $1/M$ and $1-(2M)^{-1/2}$ for the random and fully-connected
networks, respectively.

At the mean-field level, the energy reads
\begin{eqnarray}
E \approx -J_1 k^2- \frac{J_2}{M} nk^2 
   = -2J_1M\psi-2J_2\sqrt{2M\psi}.
\end{eqnarray}
To evaluate the entropic contribution to the free energy at finite temperatures,
we write the number of accessible configurations
\begin{equation}
\Omega_{\psi}=\left(\begin{array}{c}N \\ n \end{array}\right)
\left(\begin{array}{c}\frac{1}{2}n(n-1) \\ M \end{array}\right),
\end{equation}
with $n=\sqrt{2M/\psi}$.
With the help of Stirling's formula,
the entropy is expressed as a function of $\psi$:
\begin{equation}
S=\ln\Omega_{\psi} \approx
    n + n\ln\left(\frac{N}{n}\right)+M+M\ln\left(\frac{1}{\psi}-1\right),
\end{equation}
the first two terms of which are negligible in case $\psi \approx 1/M$.

\begin{figure}
\centerline{\epsfig{file=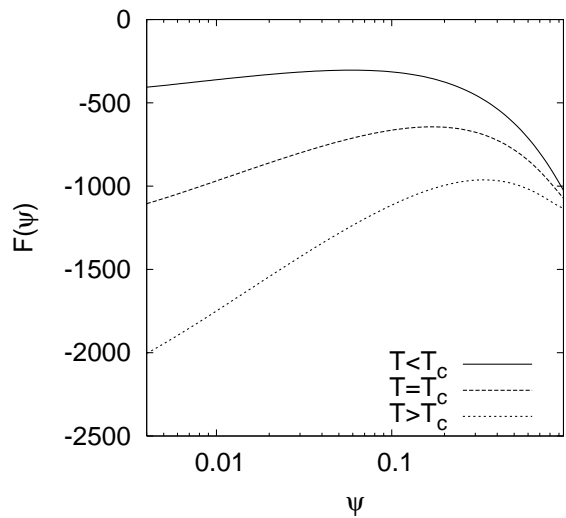,width=7cm,angle=270}}
\caption{Free energy $F(\psi)$ versus the order parameter $\psi$,
obtained from the mean-field analysis of the system of $N = 1000$ nodes and $M= 250$ links
with $J_1 = 2$ and $J_2 = 1$ at temperature $T= 0.2, 0.6$, and $1$.
At temperature $T$ higher/lower than the transition temperature $T_c = 0.6$,
the free energy of a random network is lower/higher than that of
a fully-connected network. 
At $T=T_c$, the two phases coexist and a first-order phase
transition takes place between them.} \label{fig:FreeE}
\end{figure}

Note that both the energy and the entropy are monotonically decreasing functions
of the order parameter $\psi$.  As displayed in Fig.~\ref{fig:FreeE},
the free energy reaches a maximum between $1/M$ and $1-(2M)^{-1/2}$
at finite temperatures.
This indicates that the phase transition is again first-order.
Taking the leading terms in the entropy, we compute the free
energy in the two phases:
\begin{eqnarray}
F_{F} &\equiv& F(\psi{=}1{-}(2M)^{-1/2}) 
         \approx -2J_1M-2\sqrt{2M}J_2 -T\sqrt{2M}(\ln\sqrt{2M}+1) \nonumber \\
F_{R} &\equiv& F(\psi{=}M^{-1}) 
          \approx -2J_1-2\sqrt{2}J_2 -TM(\ln M+1). 
\end{eqnarray}
At the transition temperature $T_c$, the two phases have the same
free energy (see Fig.~\ref{fig:FreeE}). Equating the above two
expressions, i.e., $F_{F}(T_c )=F_{R}(T_c )$, we obtain the
transition temperature
\begin{equation}
T_c=\frac{2J_1+2\sqrt{2/M}J_2 }{\ln M+1},
\end{equation}
which depends on $J_1$ but very weakly on $J_2$. In particular,
the transition temperature decreases logarithmically with $M$, which
indicates that in the thermodynamic limit random networks are
prevalent at finite temperatures.

\subsection{Phase transition between star and random networks}
The phase transition between a star network and a random one is
described by the order parameter~\cite{ref:Vicsek}
\begin{equation}
\Phi=\frac{k_{max}}{M},
\end{equation}
where $k_{max}$ is the largest degree of the network.
The star-network phase is characterized by a large value of the order parameter,
i.e., $\Phi \approx 1$.
In terms of this order parameter, the energy function is approximately given by
\begin{eqnarray}
E \approx -J_1 k_{max} - \frac{J_2}{M} k_{max}^2 
   = -J_1 M\Phi -J_2 M\Phi^2
\end{eqnarray}
although it does not represent the random network phase well. To
estimate roughly the transition temperature $T_c$, we consider the
number of accessible configurations
\begin{equation}
\Omega_{\Phi}=N\left(\begin{array}{c}N-1 \\ k_{max}\end{array}\right)
        \left(\begin{array}{c}\frac{1}{2}(N-1)(N-2) \\ M-k_{max} \end{array}\right).
\end{equation}
This gives the leading expression of the entropy in terms of the order parameter $\Phi$:
\begin{equation}
S = -\Phi M \ln N
\end{equation}
up to an additive constant, as reported in Ref.~\cite{ref:Vicsek}.

Comparing the resulting free energy function in the two phases,
namely, at $\Phi=1$ and at $\Phi \approx 0$, we find that the
former, corresponding to the star network, provides the global
minimum of the free energy at low temperatures. The transition
temperature below which the random network turns to a star network
is given by
\begin{equation}
 T_c \approx \frac{J_1+J_2}{\ln N} ,
\end{equation}
which again decreases logarithmically with the size $N$. It is
thus concluded that the phase transition occurs only in a finite
system.

\section{Numerical results}
\label{sec:num}

In this section we present results of Monte Carlo simulations performed
for various values of $J_1/J_2$ and system size $N=100, 200, 400$, and $1000$.
The average connectivity of the whole network is fixed to be $\langle k \rangle=0.5$,
so that the number of total links is given by $M= N\langle k \rangle /2=N/4$.
First, a random network is generated by connecting randomly selected two nodes $M$ times,
then the system is annealed from high temperatures via the standard Metropolis algorithm
with randomly selected links rewired. We allow a dangling node to be deprived of its link
and expect many isolated nodes to appear at the transition. 
Typical network configurations obtained thus are shown in Fig. \ref{fig:net}.

\begin{figure}
\centerline{\epsfig{file=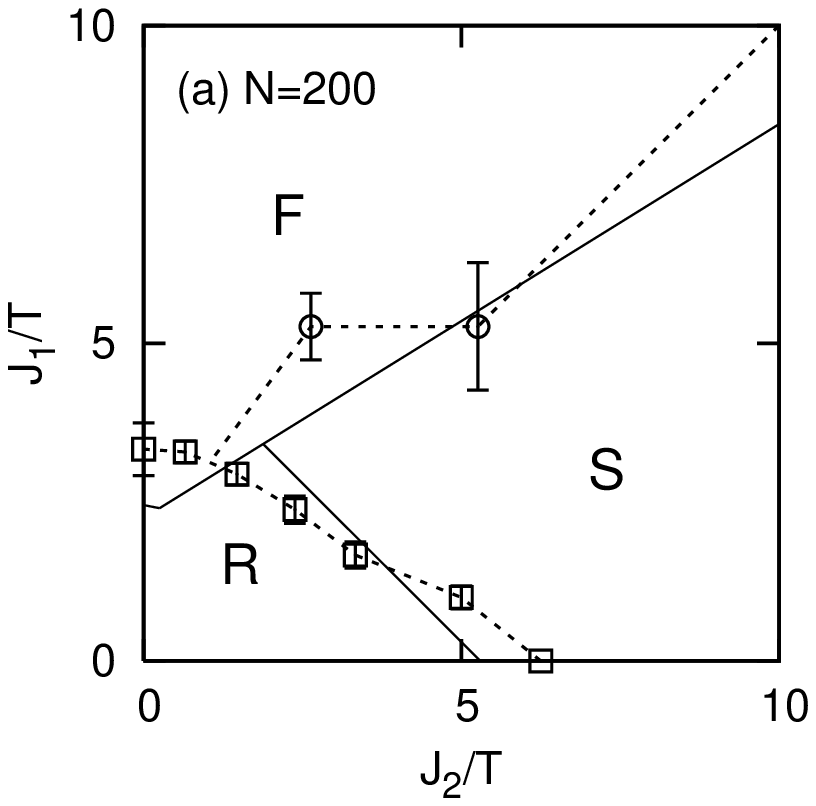,width=7cm}}
\centerline{\epsfig{file=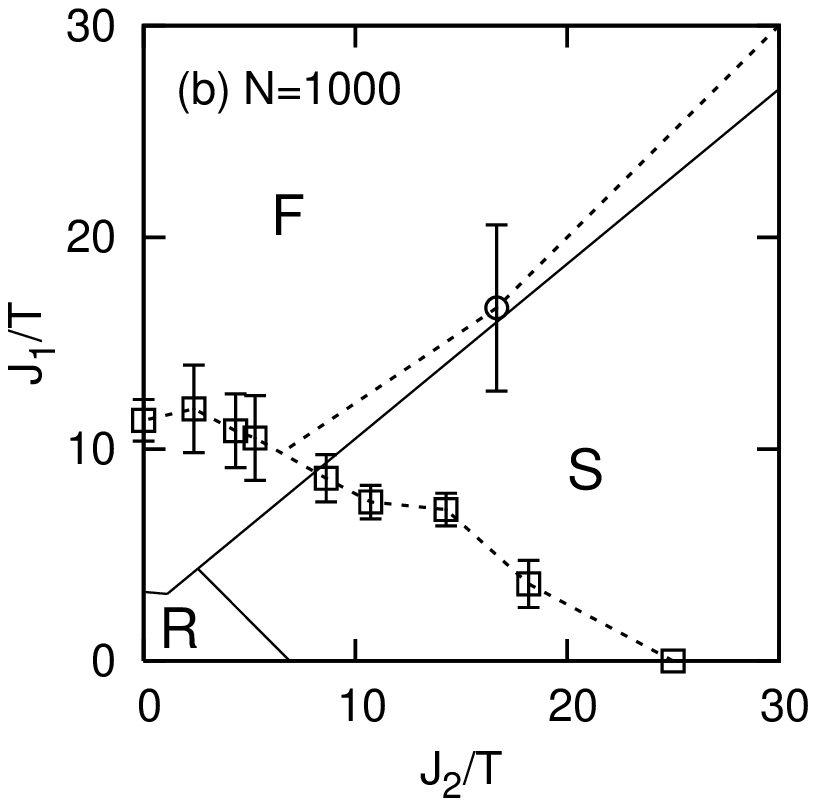,width=7cm}}
\caption{Topological phase diagram
on the $(J_1/T, J_2/T)$ plane for size $N = $(a) $200$ and (b) $1000$.
Simulation results, obtained via cooling from a random network, are depicted by symbols,
separating regions of random (R), fully-connected (F), and star (S) networks.
Dashed lines are merely guides to the eye, and in particular those
at low temperatures, i.e., boundaries between F and S regions, are speculative and
just represent $J_1/J_2 =1$,
which is the minimum value to produce a fully-connected network as the ground state in
numerical simulations.
Also shown are results from the mean-field analysis, plotted by solid lines.
}
\label{fig:PD}
\end{figure}

Figure~\ref{fig:PD} exhibits the phase diagram of networks
constructed by the energy function given by Eq. (1) for $N=200$
and $1000$. It is observed in both numerical and mean-field
results that phase boundaries depend on the system size. As
predicted in mean-field analysis, the system undergoes a
discontinuous topological phase transition from a random network
to a compact one such as a star or a fully-connected network as
the temperature is lowered.
When $J_1$ is smaller/greater than $J_2$, the ground state is given by the star/fully-connected network.
For $J_1/J_2\gtrsim 1$, in particular, the star-network phase emerges as an intermediate state,
so that there occur double transitions as the temperature is lowered.
This is more evident for small system size, which is consistent with the mean-field results.

In the transition from a random to a fully-connected network of large size,
it is observed that at the instant of sudden change of topology,
the system tends to be  trapped in a characteristic multi-star network
which consists of two types of node: A few star nodes connected to all nodes and the remaining
peripheral nodes connected only to the star nodes identically.
Such a multi-star network appears during most of simulations for $J_1/J_2>1$
and $N=1000$, the largest system size considered here.
However, it disappears if more MC steps are performed, especially on a system of smaller size;
this indicates the multi-star network to be a metastable state.
Although such a metastable state has also been considered in the mean-field analysis,
our free energy function does not have any local minimum corresponding to the metastable state.

Shown in Fig.~\ref{fig:PD} is discrepancy between numerical and mean-field results, which
apparently grows with the system size.  Here it should be noted that our numerical results
have been obtained from cooling simulations.
Considering that the phase transition is strongly discontinuous, accompanied with the hysteresis,
we expect to obtain a higher transition temperature in heating simulations.
With $T_1$ and $T_2$ denoting the transition temperatures in cooling and heating simulations,
respectively, Fig.~\ref{fig:PD} displays that the annealed transition temperature $T_1$ is lower
than the mean-field transition temperature $T_{MF}$.
In the specific case of  the transition from a random network to a fully-connected one,
$T_1$ in numerical simulations may be estimated as follows:
At temperature $T=T_1$, we assume that the slope of the free energy vanishes
at $\psi=1/M$ (corresponding to the random network phase) and write
$0=dF/d\psi | _{\psi=1/M} \approx -2J_1M-\sqrt{2}J_2M+T_1 M^2$.
This leads to the transition temperature in annealing
\begin{equation}
T_1= \frac{2J_1+\sqrt{2}J_2}{M},
\end{equation}
which decreases with the size much faster than the mean-field transition temperature.
The discrepancy growing with the system size is thus explained.
Similarly, at $T=T_2$ in heating from a fully-connected network, we write
$0=dF/d\psi | _{\psi\approx 1} \approx -2J_1M-J_2\sqrt{2M}
    + T_2\sqrt{M/2} \ln(\alpha \sqrt{M/2})$,
which gives the transition temperature $T_2$ increasing with the system size,
\begin{equation}
T_2= \frac{4\sqrt{2M}J_1+4J_2}{\ln M +2\ln\alpha -\ln 2}.
\end{equation}
As a result, the hysteresis becomes more evident as the system
size is increased in simulations. Therefore, when the system size
is sufficiently large, $T_{MF}$ is expected to locate between
$T_1$ and $T_2$.

\begin{figure}
\centerline{\epsfig{file=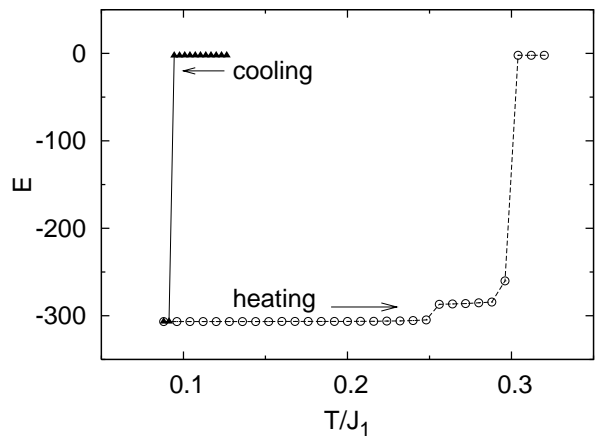,width=8.5cm}}
\caption{Hysteresis in the energy versus temperature  curves in cooling and heating simulations
for $J_1/J_2=2.5$ and $N=1000$.
}
\label{fig:hysterisis}
\end{figure}

We display in Fig.~\ref{fig:hysterisis} the behavior of the energy for $J_1/J_2=2.5$ and $N=1000$,
as the temperature is lowered (cooling from the random network) or raised (heating
from the fully-connected network which is the ground state).
Here the hysteresis is manifested.
Note, however,  that $T_2$ obtained numerically is lower than $T_{MF}$, unlike
the above conjecture.
We presume that this inconsistency results from fluctuations neglected in estimating $T_2$.
Indeed fluctuations should assist the system to overcome the free energy barrier, thus
lowering $T_2$ and also suppressing its indefinite increase with the size.

\section{Conclusion}
\label{sec:conclusion}

We have constructed equilibrium networks by means of a topological
energy function which depends quadratically on the node degrees.
The topological phase transitions between random, star, and fully-connected
networks have been studied both analytically and numerically, through the use of
mean-field and simulation methods.
It has been observed that the system undergoes discontinuous transitions
between the three types of network as the temperature or the interaction strength
relative to the node term is varied.  Here the transition temperatures in general
decrease logarithmically as the system size grows.
The quantitative discrepancy between the mean-field and simulation results
is attributed to the marked hysteresis associated with the strong first-order nature
of the transition.

\ack

We acknowledge the support from the MOST/KOSEF through the National Core Research Centre 
for Systems Bio-Dynamics (R15-2004-033) and from the MOE through the BK21 Program.

\end{document}